\documentclass[a4paper,fleqn,usenatbib]{mnras}

\pdfoutput=1
\usepackage[T1]{fontenc}
\usepackage{ae,aecompl}
\usepackage{float}
\usepackage{cleveref}
\crefname{section}{§}{§§}
\Crefname{section}{§}{§§}

\usepackage{graphicx}	
\usepackage{amsmath}	
\usepackage{amssymb}	

\title[Velocity Dispersion Profiles of Clusters]{The Shape of Velocity Dispersion Profiles and the Dynamical State of Galaxy Clusters}

\author[Costa, Ribeiro \textit{\&} de Carvalho]{
A. P. Costa$^{1}$\thanks{E-mail: apcosta@uesc.br},
A. L. B. Ribeiro$^{1}$\thanks{E-mail: albr@uesc.br} and 
R. R. de Carvalho$^{2}$\thanks{E-mail: rrdecarvalho2008@gmail.com}
\\
$^{1}$Laborat\'orio de Astrof\'{\i}sica Te\'orica e Observacional, Universidade Estadual de Santa Cruz-45650-000, Ilh\'eus-BA, Brazil\\
$^{2}$Divis\~ao de Astrof\'{\i}sica (INPE-MCT), S\~ao Jos\'e dos Campos, 12227-010, SP, Brazil
}

\date{Accepted 2017 September 22. Received 2017 September 17; in original form 2017 August 30}

\pubyear{2017}

\begin{document}
\label{firstpage}
\pagerange{\pageref{firstpage}--\pageref{lastpage}}
\maketitle

\begin{abstract}
Motivated by the existence of the relationship between the dynamical state of clusters and the shape of the velocity dispersion profiles (VDP), 
we study the VDPs for Gaussian (G) and Non-Gaussian (NG) systems for a subsample of clusters from the Yang catalog. The groups cover a redshift interval of 
$0.03\leq z\leq0.1$ with halo mass $\geq 10^{14}$M$_{\odot}$. We use a robust statistical method, Hellinger Distance, to classify the dynamical state of the systems 
according to their velocity distribution. The stacked VDP of each class, G and NG, is then determined using either Bright or Faint galaxies. The stacked VDP for G groups 
displays a central peak followed by a monotonically decreasing trend which indicates a predominance of radial orbits, with the Bright stacked VDP showing lower velocity dispersions in all radii. 
The distinct features we find in NG systems are manifested not only by the characteristic shape of VDP, with a depression in the central region, but also by a possible higher infall rate associated 
with galaxies in the Faint stacked VDP.

\end{abstract}




\section{Introduction}

In a hierarchical universe ($\Lambda$CDM), galaxy clusters constitute the last representative blocks of the large scale structures to be formed by accretion of lower mass systems from the general field and filamentary regions (e.g. 
\citealt{1972ApJ...176....1G}; \citealt{1974ApJ...187..425P}; \citealt{1991ApJ...379...52W}; \citealt{2004ApJ...609L..49E}). In this scenario, clusters represent the most massive structures recently collapsed in the universe and also the most dynamically ''immature'' (e.g. \citealt{boylan2009resolving}; \citealt{2010arXiv1005.2502I}). An important tracer of the dynamical state of galaxy clusters, is the velocity dispersion profile (e.g. \citealt{1979AJ.....84...27S}), which provides valuable information on the degree of anisotropy of galaxy orbits and is also related to the cluster density profile (e.g. \citealt{1996MNRAS.278..321J}; \citealt{1997A&A...321...84B}; \citealt{1998A&A...331..439A}). \cite{1996ApJ...472...46M}, after numerically integrating the Boltzmann-Liouville equation for colliding galaxies in the potential well of clusters, find that centrally increasing or decreasing VDPs are associated to the balance between galaxy interactions and the shape of the dark matter distribution. Using the CNOC1 survey, \cite{1997ApJ...476L...7C} show that velocity dispersion rises from 0.1 virial radius, reaches a peak around 0.3 virial radius and then presents a roughly flat profile with a very slight decline (see also \citealt{2000AJ....119.2038V} ). With a different approach, studying the kinematics of groups in the SDSS DR7, \cite{2012ApJ...758...50L}  find that the average velocity dispersion within the virial radius is a strongly increasing function of the central galaxy mass. Generally, increasing or decreasing features of VDPs could be related to the influence of cD galaxies in the first radial bin, two-body relaxation, and orbit circularization in the central region of clusters (\citealt{1996MNRAS.279..349D}; \citealt{1998ApJ...505...74G} ). In other words, a critical factor influencing the shape of the VDPs is the dynamical state of clusters (e.g. \citealt{1996MNRAS.279..349D}; \citealt{2009ApJ...702.1199H}; \citealt{2014MNRAS.438.3049P}).

Although galaxy clusters may go through different dynamical states, they are usually classified  in only two classes:
relaxed and unrelaxed (e.g. \citealt{2009ApJ...702.1199H}; \citealt{2011MNRAS.413L..81R}; \citealt{2017MNRAS.464.2502C}). Dynamically relaxed clusters are expected to have Gaussian line-of-sight velocity distributions (e.g. \citealt{1977ApJ...214..347Y}; \citealt{1993AJ....105.1596B}; \citealt{2013MNRAS.434..784R}), while unrelaxed systems present significant departures from the Gaussian distribution, which can be seen as evidence of different on-going physical processes: presence of interlopers; displacement of the brightest cluster galaxy (BCG) from the peak of the projected galaxy density, or from the peak of the X-ray emission; circular orbits; and the most frequent, the presence of substructures (e.g. \citealt{1983ApJ...274..491B}; \citealt{2004ApJ...617..879L}; \citealt{2009ApJ...693..901O}; \citealt{2014ApJ...797...82L}; \citealt{2016MNRAS.457.4515R}). All these effects reinforce the idea that the shape of the VDPs can be assessed by investigating the dynamical state of galaxy clusters. \cite{2009ApJ...702.1199H}, using a sample selected from CNOC2, find that the VDPs of groups with non-Gaussian velocity distribution are significantly different from the Gaussian ones. Also, \cite{2012MNRAS.421.3594H} studying groups of intermediate redshift from the GEEC (Group Environment and Evolution Collaboration) catalog, 
show that there is a relationship between the shape of the VDPs and the presence of substructures in clusters. Using the methodology prescribed by \cite{2006A&A...448..155B} to compute the VDPs, they notice that all groups with substructures have strictly increasing VDPs. 

All these studies show not only the importance of the use of VDPs as a proxy of the dynamical state of clusters, but also point to a possible relationship, between their behavior and the different physical mechanisms in action within clusters. In previous studies, the shape of VDPs was analyzed considering a number of dynamical indicators (e.g. \citealt{1996ApJ...472...46M}; \citealt{2009ApJ...702.1199H}). In this Letter, we study the VDPs of 177 groups selected by \cite{1538-3881-154-3-96} from the Yang catalog (\citealt{2007ApJ...671..153Y}), taking into account the dynamical state of the groups based on a robust statistical method ($\S$ \ref{sec22}). We describe the sample used in Section 2. The results obtained for the VDPs are presented in Section 3 followed by a discussion in Section 4. Throughout this work we assume a $\Lambda$CDM cosmology with $\Omega_{M}=0.3$,  $\Omega_{\Lambda}=0.7$ and H$_{0}= 100$ km s$^{-1}$ Mpc$^{-1}$.

\section{Data and Methodology}
\label{sec2}
In this section we will briefly describe the sample and the procedures used to identify the dynamical state of the groups. For more details on the sample and dynamical classification, see \cite{1538-3881-154-3-96}.

\subsection{Yang Groups}
To study the VDPs of Yang groups, we use an updated sample of Yang catalog (\citealt{2007ApJ...671..153Y}) presented by \cite{1538-3881-154-3-96} based on  593,736 galaxies from SDSS-DR7, supplemented with additional 3,115 galaxies with redshifts from different sources. The galaxies from SDSS-DR7 covers an interval of $0.03 \leq z \leq 0.1$ and r magnitudes brighter than 17.78 (spectroscopic completeness limit of the survey), ensuring that we cover the luminosity function up to $M^{*}+1$ for all systems. The membership, $R_{200}$, $M_{200}$ and velocity dispersion for each group were re-estimated by shift-gapper technique and virial analysis following prescription described in \cite{2009MNRAS.392..135L}. Finally, only systems with more than 20 galaxies within $R_{200}$ are used. After these constraints in redshift ($0.03\leq z\leq0.1$) and richness, the number of groups/clusters remaining is 319.

\subsection{Classifying the dynamical state of groups}
\label{sec22}

To classify the dynamical state of the 319 groups/clusters we use a new method, Hellinger Distance (HD), which
is based on their line-of-sight velocity distribution. Succinctly, HD (\citealt{le2012asymptotics}) measures how far from
a gaussian a given distribution is. It was first introduced in astronomy by \cite{2013MNRAS.434..784R}, studying the degree of gaussianity of the velocity distribution of galaxies in Berlind's groups (\citealt{2006ApJS..167....1B}). 
We estimate HD using codes available in R environment under the distrEx package (\citealt{Ruckdeschel2006}). Only Gaussian (G) or Non-Gaussian (NG) systems with reliability greater than 70$\%$ were considered (see de Carvalho et al. 2017 for details). Also,  we find that, through the relation between $M_{200}$ and $N_{R200}$ (where $N_{R200}$ is the number of galaxies within $R_{200}$ with $M_{r}\leq-20.55$), a mass cutoff of $10^{14}M_{\odot}$ corresponds to $N_{R200}=20$. These constraints define a final sample of 177 groups/clusters, being 143 G and 34 NG. Also, in the analysis that follows, we consider two specific domains of luminosity for the final sample: Bright, using galaxies with $M_{r}\leq-20.55$, probing the systems up to $M^{*}+1$ (\citealt{2001AJ....121.2358B}); Faint, using galaxies with $-20.55<M_{r}\leq -18.40$. The faint domain is only analyzed for groups/clusters in the range $0.03\leq z \leq0.04$, examining the luminosity function down to $\sim M^{*}+3$.

\section{Velocity Dispersion Profiles Analysis}
\label{sec3}

The study of VDPs is a powerful tool for doing dynamical analysis of galaxy clusters (e.g. \citealt{1979AJ.....84...27S}; \citealt{1996MNRAS.279..349D}; \citealt{2014MNRAS.438.3049P}). We probe the VDPs for clusters classified as G and NG, first using all galaxies and then exploring both luminosity domains previously defined. We estimate  the cumulative velocity dispersion with 
the robust bi-weight scale estimator described by \cite{1990AJ....100...32B}, instead of the methodology presented by \cite{2006A&A...448..155B}. This method requires a careful choice of the kernel scale parameter which is very sensitive, namely, for large values some VDP features may disappear, while choosing small values tend to add fake components to VDPs. Different choices of the kernel parameter and
their effects are detailed in \cite{2009ApJ...702.1199H} and \cite{2014MNRAS.438.3049P}.

Subsequently, we build composite clusters, since this is the most appropriate way to investigate galaxies 
in multiple galaxy systems (\citealt{2003ApJ...585..205B}; \citealt{2010MNRAS.409L.124R}). Also, by using composite clusters we reduce asymmetries in the galaxy distribution (\citealt{biviano2001tracing}).
We create two composite groups, G (composed of 143 systems) and NG (composed of 34 systems). The stacked VDPs (SVDPs) for these two classes have distances to the group centre normalized by $R_{200}$ and their peculiar velocities are  scaled to the cluster velocity dispersion. VDPs are obtained by ordering the members of the stacked cluster in distance and measuring the scale within the 
radius of each galaxy considered in each step starting with the first ten galaxies ordered in distance, since this number of galaxies corresponds to an efficiency between 70$\%$ and 80$\%$ of the estimates obtained by the bi-weight estimator (\citealt{1990AJ....100...32B}) and are also sufficient to obtain an unbiased estimate of a cluster dispersion (\citealt{2006A&A...456...23B}).
Fiber collision problem can be an issue when measuring galaxy clustering statistics on small scales and for that reason we estimate that the radius within which the effect can be important is $0.15 h^{-1}$Mpc ($55^{\prime\prime}$). With a typical $R_{200}$ (evaluated from the 319 clusters) of $0.95 h^{-1}$Mpc, we distrust measurements within $R_{p}/R_{200}\sim0.16$.

\begin{figure} 
\hspace*{-0.4in}
       \includegraphics[width=9.3cm, height=9.3cm]{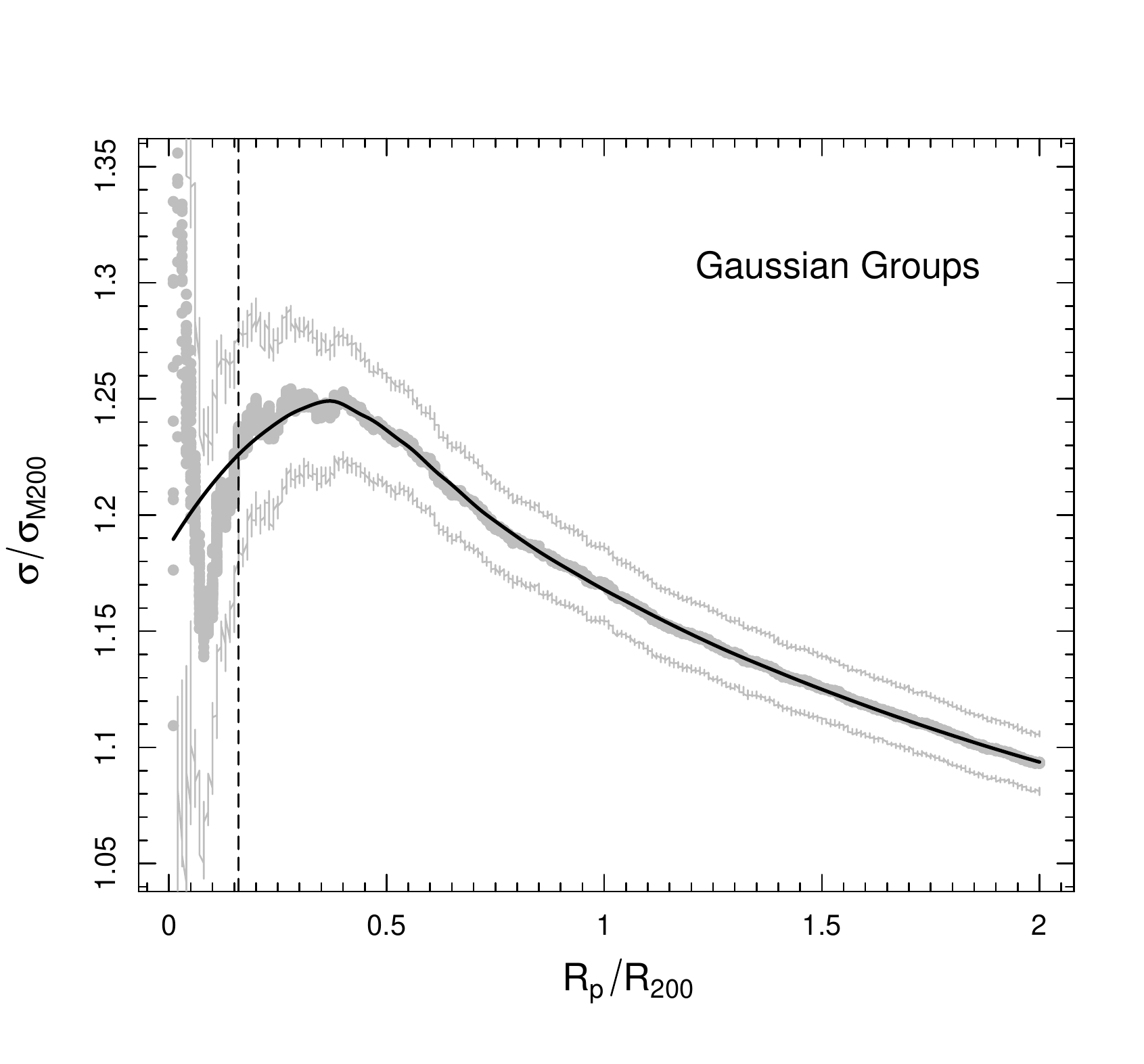}
      \caption{SVDP for clusters classified as G. The grey points represents the velocity dispersion estimated in each $R_{p}/R_{200}$ considered and the solid black line 
      represents the best LOWESS estimated to the all galaxies along the profile. The grey lines are representative of the confidence intervals (90$\%$) obtained through 1000 bootstraps, and the 
      vertical dashed line represents the inner radius to which the fiber collision problem can be considered.}
    \label{fig01}
    \end{figure}

\begin{figure} 
\hspace*{-0.2in}
       \includegraphics[width=9.3cm, height=9.3cm]{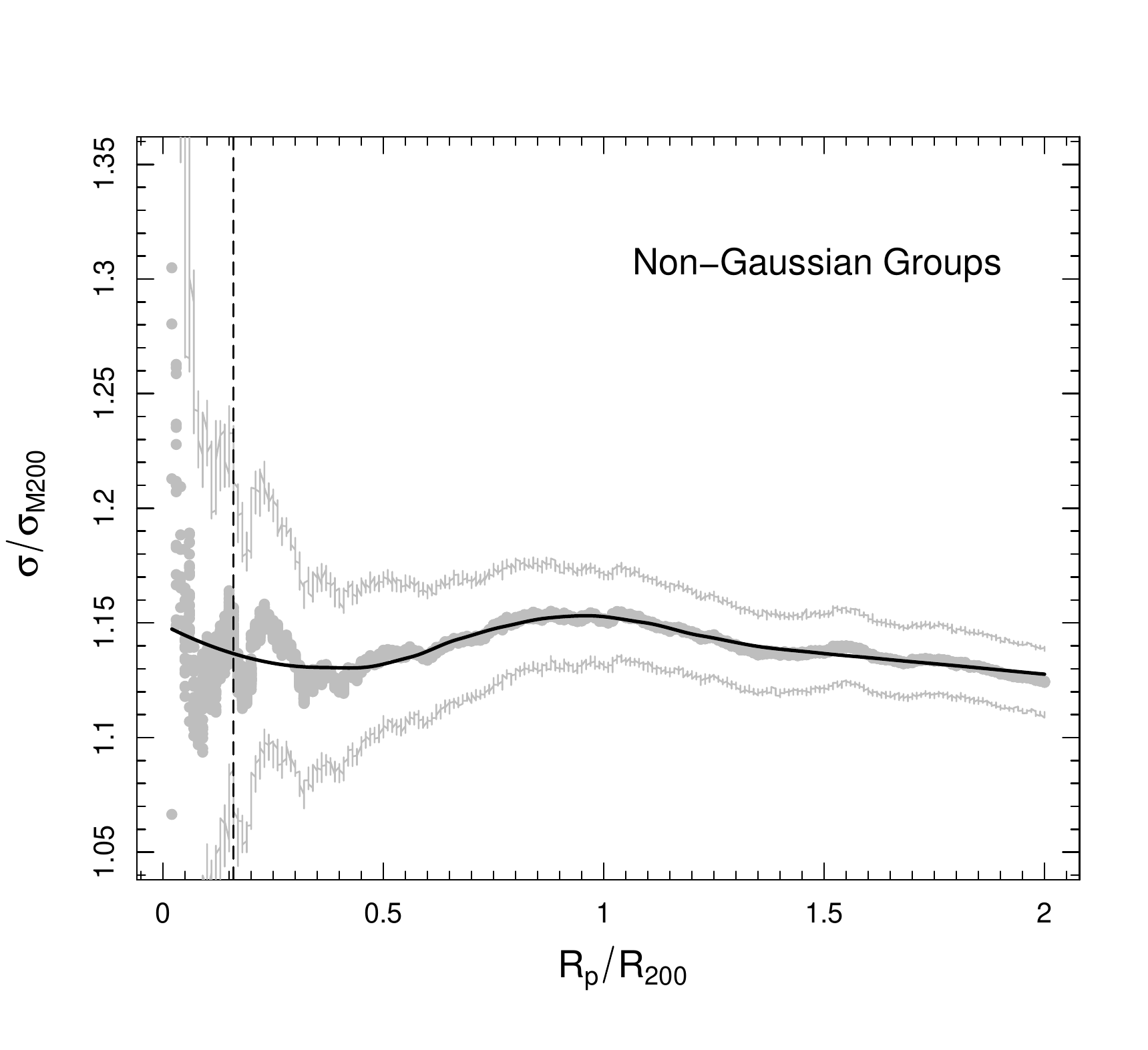}
      \caption{SVDP for clusters classified as NG. The points and lines (grey and black) have the same meaning as those explained in Fig.\ref{fig01}. }
    \label{fig02}
    \end{figure}

In Figure \ref{fig01}, we plot the SVDPs for G groups, $\sigma/\sigma_{M200}$, as a function of $R_{p}/R_{200}$, where $\sigma_{M200}$ represents the estimated velocity dispersion measured by the shift-gapper technique \citep{2009MNRAS.392..135L}. From this figure, we clearly see that velocity dispersion
increases to approximately $R_{p}/R_{200} \sim 0.35$, and for larger radii the profile is monotonically decreasing. For NG clusters, Fig.\ref{fig02}, we see that the shape of the SVDP exhibits a central depression and then increases from $R_{p}/R_{200} \sim 0.5$ to 1.0. This upward trend is also observed in two NG groups studied by \cite{2009ApJ...702.1199H}, which is interpreted as a signature of merge by  \cite{1996ApJ...472...46M}.

These differences between SVDPs for G and NG groups, make it more evident that these systems are in different stages of evolution. In the SVDP for G groups, we see an increase up to $R_{p}/R_{200}\sim 0.35$, which might indicate a cool core remnant, characteristic of relaxed systems, originating from a process of violent relaxation  as noticed by \cite{1988AJ.....95..985D} and \cite{1996MNRAS.279..349D}. 
This tendency for relaxed systems is also consistent with the results presented by \cite{1998ApJ...505...74G} and \cite{2017arXiv170808541C}, who find similar behavior for a stacked sample of clusters classified as regular using the Dressler-Schetman test (\citealt{1988AJ.....95..985D}).
In regions external to $R_{p}/R_{200} \sim 0.35$, the SVDP decreases, as expected for G systems (without substructures), i.e., in regions where there is a predominance of radial orbits (e.g. \citealt{1997MNRAS.286..329N}). 
As for the SVDP of NG systems, the depression seen in the central region, internal to $R_{p}/R_{200}\sim 1.0$, suggests that subgroups of galaxies may be in some stage of merging, a scenario consistent with the absence of a cool dense core, which allows substructures penetrate deep inside the potential well and disturb the cluster central dynamics (\citealt{2003ApJ...590..225C}; \citealt{2011MNRAS.413L..81R}). Finally, for regions with $R_{p}/R_{200}$ $\gtrsim$ 1.0, the SVDP becomes slightly flat, 
suggesting that the measured velocity dispersion at larger radii is more representative of the total kinetic energy of the cluster galaxies (e.g. \citealt{1996ApJ...473..670F}; \citealt{2003ApJ...585..205B}; \citealt{2011A&A...526A.105Z}).

\begin{figure} 
\hspace*{-0.2in}
       \includegraphics[width=9.3cm, height=9.3cm]{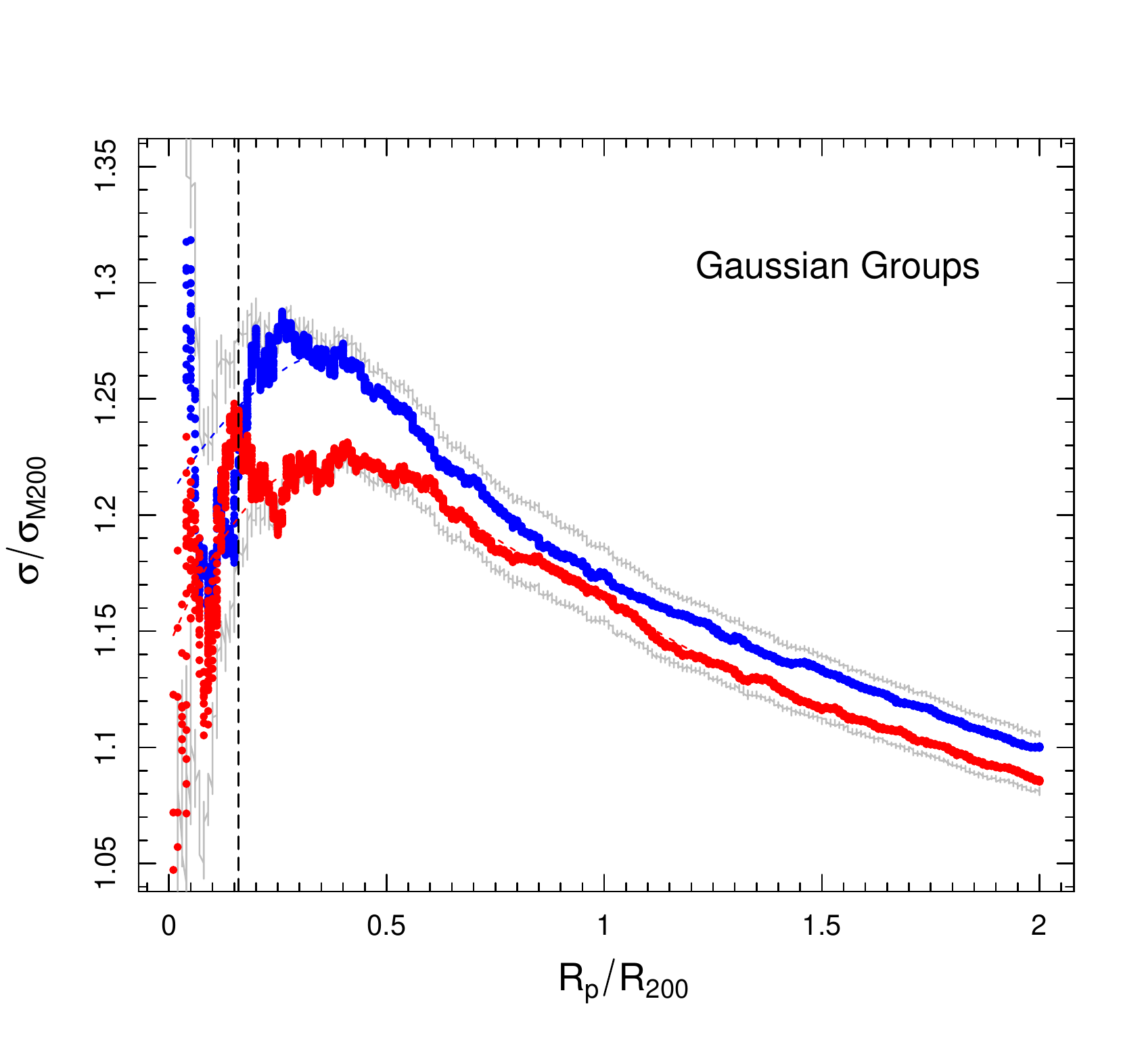}
      \caption{VDP for G systems with superposition of profiles for galaxies classified as Bright(red) and Faint(blue). 
      Both profiles are within the confidence envelope obtained through all objects in this sample.}
    \label{fig03}
    \end{figure}

\begin{figure} 
\hspace*{-0.2in}
       \includegraphics[width=9.3cm, height=9.3cm]{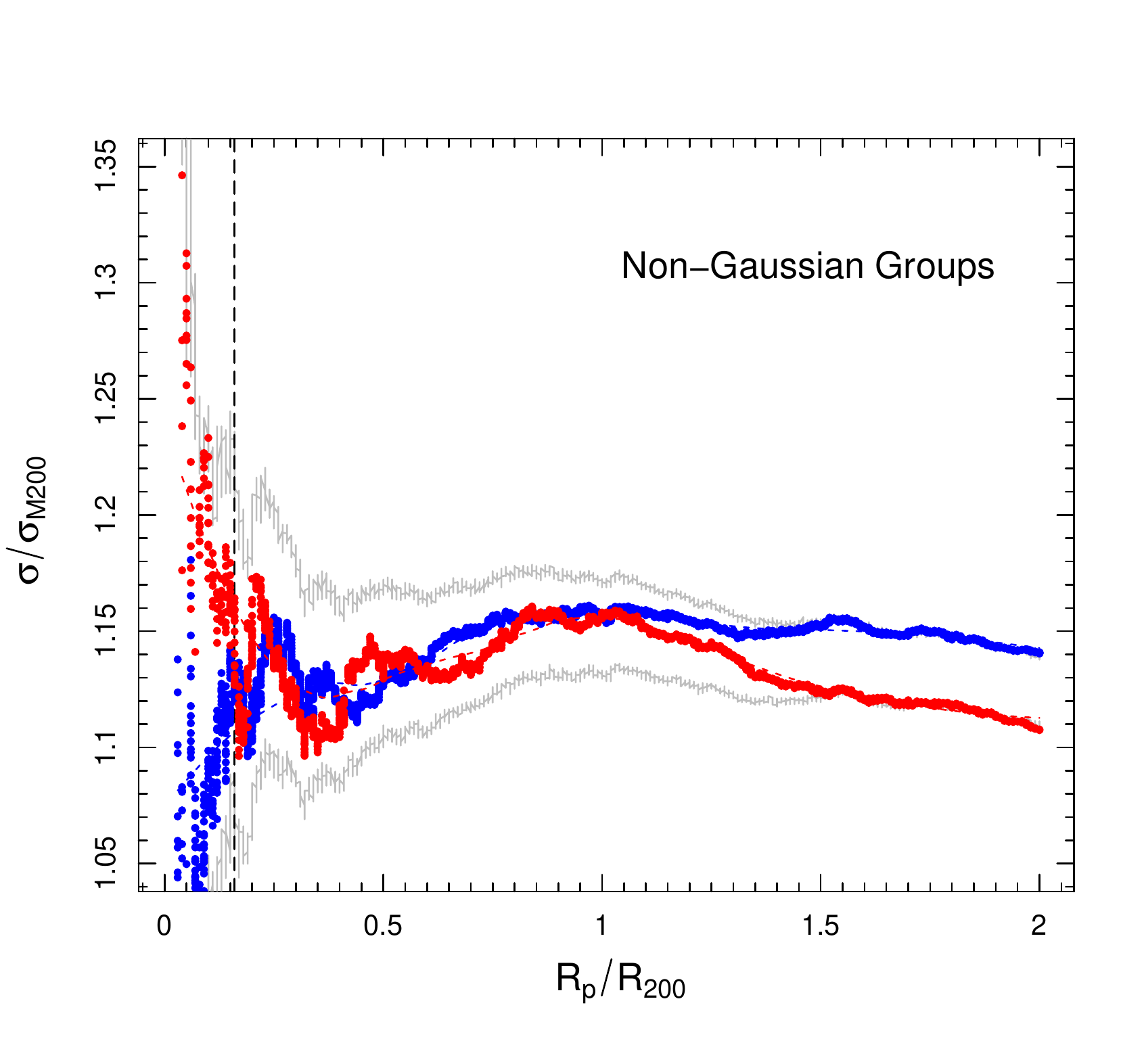}
      \caption{VDPs for the Bright and Faint populations belonging to the NG sample. The profiles blue and red represents the two domains of luminosity considered. }
    \label{fig04}
    \end{figure}

We extend our analysis of the SVDPs to the two luminosity domains, Bright and Faint, and two dynamical classes, G and NG as displayed in Figures \ref{fig03} and \ref{fig04}. For reference, we superimpose the SVDPs obtained when all galaxies are used. The red and blue profiles represent the SVDPs measured using only the Bright and Faint galaxies, respectively. Both profiles are inside the confidence envelope for G and NG groups, indicating that their characteristics do not reflect unphysical effects.

In Fig.\ref{fig03}, we see that the Bright and Faint SVDPs follow the same general trend: the Bright component exhibits a lower velocity dispersion {\it wrt} the Faint one along the entire profile, as also obtained by \cite{goto2005velocity} ($M_{z} < -23.0$) and for bright objects ($M_{R} \leq -21.5$) in G groups by \cite{2010MNRAS.409L.124R}. Results from \cite{2007A&A...471...17A}, considering the brightest galaxies ($M_{r} < -21.0$) for their sample of 88 nearby galaxy clusters from SDSS-DR4, seem to corroborate this trait. 
In Fig. \ref{fig04}, we notice that up to $R_{p}/R_{200} \sim 1.0$ the SVDP is not only considerably reduced {\it wrt} what we find for G systems but also seem to be indistinguishable for both luminosity regimes. This behavior reinforces the fact that NG clusters are dynamically distinct and less evolved than the G ones. In this case, we do not see any significant segregation in the inner region as we found for G clusters. However, for  $R_{p}/R_{200} > $1.0, the Faint SVDP predominates over the Bright one. A possible interpretation for this effect is that the Faint SVDP may be manifesting a larger number of infalling galaxies (e.g. \citealt{2000ApJ...539..561C}). This view agrees with the results of \cite{2011MNRAS.416.2882M} who have shown that within 1.0 - 2.0Mpc, infalling galaxies are the dominant population in phase space for certain values of line-of-sight velocities. In fact, these findings are in agreement with those encountered by other authors (e.g., \citealt{1996ApJ...470..724M}; \citealt{1999ApJ...518...69M}; \citealt{2001ApJ...547..609E}) who have demonstrated that groups infalling into clusters are dominated by blue, emission line galaxies with larger velocity dispersion than the red and more evolved galaxies.
  
\section{Discussion}

In this Letter, we study the VDPs for a sample of 177 galaxy clusters from the Yang catalog (\citealt{2007ApJ...671..153Y}), with the main
objective of finding a relation between the shape of the VDPs and the dynamical state of clusters. The dynamical state of each cluster (G or NG) is defined following a robust statistical method (Hellinger Distance), fully described in \cite{1538-3881-154-3-96}. From our analysis, important differences emerge when we compare the SVDPs behavior in G and NG groups. 

\begin{enumerate}
 \item First, the shape of the SVDP of G groups
shows the characteristic behavior of quiescent systems,
with a peak corresponding to the central regions that already underwent violent relaxation,
and now present cool cores with galaxies in isotropic orbits following a Gaussian distribution function (e.g. \citealt{1967MNRAS.136..101L}; \citealt{1996grdy.conf..121W}). The peak itself and the small scale of the core (up to $R_{p}/R_{200}$ $\sim$ 0.35)
may have evolved from interactions between galaxies, or because of adiabatic compressions caused by the accumulation of infalling matter, processes that can make the core radius be diminished with time, and the central velocity dispersion be increased (\citealt{1990ApJ...359..257M}).

 \item The core of G systems is surrounded by objects in predominantly radial orbits, with decreasing velocity dispersions,
indicating the existence of accreting galaxies from the cluster outskirts (e.g. \citealt{2001ApJ...548...97S}). This general behaviour (peak + monotonic decreasing) is consistent, for example, with the work of
\cite{1988AJ.....95..985D} for Abell 1983 and DC 0428-53, clusters not showing significant substructures when the Dressler-Shectman test is applied. 
Equivalent results are also found by others authors: \cite{1996ApJ...470..724M} studying Abell 576, a galaxy cluster with a cold core, presenting a peak associated with the velocity dispersion of the nonemission line galaxy sample, and the decreasing behavior is related to the dispersion of the emission line galaxies
(in this study the nonemission line galaxy sample is significantly brighter than the emission line sample);
\cite{2003AJ....126.2152R} for Abell 496, a cluster that has a very symmetric X-ray emission, considered as relaxed in both X-rays and in the galaxy distribution by 
\cite{2000A&A...356..815D}; and \cite{2004A&A...424..779B} when estimating the velocity dispersion for the brightest elliptical galaxies from the ENACS survey.

 \item Other important feature for the SVDP of G groups is the kinematical segregation between 
the Bright and Faint components, with the Bright stacked
VDP showing lower velocity dispersions in all radii. This could be indicating that, on average,
the Faint VDP is built with more recently accreted galaxies than the Bright VDP. Some authors find similar
results for samples defined in terms of morphology or colour, see for instance \cite{goto2005velocity} and \cite{2007A&A...471...17A}.

\item Finally, for the NG groups, the
irregular trend of the inner SVDP up to $R_{p}/R_{200}$ $\sim$ 1.0, exhibiting a depression, corroborates the idea that NG systems
have different VDPs when compared to Gs, possibly due to
mergers and infall of subgroups in the central region. Also,
the excess of the Faint SVDP over the Bright one in the
outskirts of the systems, suggests a higher infall rate, probably dominated by faint objects (e.g. \citealt{2005AJ....130.1482R}; \citealt{2015ApJ...806..101H}), onto NGs. 

\end{enumerate}

\section*{Acknowledgements}

APC thanks CAPES financial support, ALBR thanks CNPq, grant 309255/2013-9 and RRdC acknowledges financial support from FAPESP through grant $\#$2014/11156-4.
We also would like to thank the referee, Andrea Biviano, for comments and suggestions that helped improving the manuscript.




\bibliographystyle{mnras}
\bibliography{references.bib} 








\bsp	
\label{lastpage}
\end{document}